\documentstyle[12pt, hyperref]{article}

\catcode`\@=11
\def\theequation{\thesubsection.\arabic{equation}}
\@addtoreset{footnote}{section}
\@addtoreset{footnote}{subsection}
\@addtoreset{equation}{subsection}
\catcode`\@=12
\catcode`\@=11

\newcounter{app}

\def\app{\setcounter{equation}{0}
\def\theequation{A\arabic{app}.\arabic{equation}}\par
   \addvspace{4ex}
   \@afterindentfalse
  \secdef\@app\@dapp}
\newcommand\@app{\@startsection {app}{1}{0ex}%
                                   {-3.5ex \@plus -1ex \@minus -.2ex}%
                                   {2.3ex \@plus.2ex}%
                                   {\normalfont\Large\bf}}

\def\@dapp#1{%
{\parindent \z@ \raggedright  \bf #1}\par\nobreak}
\def\l@app#1#2{\ifnum \c@tocdepth >\z@
    \addpenalty\@secpenalty
    \addvspace{1.0em \@plus\p@}%
    \setlength\@tempdima{8.5em}%
    \begingroup
      \parindent \z@ \rightskip \@pnumwidth
      \parfillskip -\@pnumwidth
      \leavevmode \bfseries
      \advance\leftskip\@tempdima
      \hskip -\leftskip
      #1\nobreak\hfil \nobreak\hb@xt@\@pnumwidth{\hss #2}\par
    \endgroup\fi}
\catcode`\@=12

\small  \oddsidemargin -10mm  \evensidemargin 5mm  \topmargin -10mm
\def\be{\begin{equation}}
\def\ee{\end{equation}}
\def\g{\gamma}

\renewcommand{\l}{\langle}
\renewcommand{\r}{\rangle}

\newtheorem{prop}{Proposition}
\newtheorem{remark}{Remark}

\def\bprop{\begin{prop}}
\def\eprop{\end{prop}}
\def\bremark{\begin{remark}}
\def\eremark{\end{remark}}
\begin{document}
\begin{center}
{\large\bf Hypergeometric functions
 related to Schur Q-polynomials and BKP equation}\\
\vspace{.5cm}
%\vspace{1cm}
{\bf A. Yu. Orlov\footnote{E-mail: orlovs@wave.sio.rssi.ru}}
\vspace{.2cm}\\
 Nonlinear wave processes laboratory,\\
Oceanology Institute, 36 Nakhimovskii prospekt\\
Moscow 117851, Russia.

\end{center}

%\vspace{1.5cm}
\begin{abstract}
We introduce hypergeometric functions related to projective Schur
functions $Q_{\lambda}$ and describe their properties. Linear
equations, integral representations and Pfaffian representations
are obtained. These hypergeometric functions are vacuum
expectations of free fermion fields, and thus these functions are
tau functions of the so-called BKP hierarchy of integrable
equations.
\end{abstract}

\section{Hypergeometric series}
BKP hierarchy was invented in \cite{DJKM},\cite{DJKM'} as a
certain generalization of the KP hierarchy of integrable equations
\cite{ZSh},\cite{DJKM}. In \cite{You} (see also \cite{Nimmo}) the
role of  projective Schur functions \cite{Mac} as rational
solutions of the BKP hierarchy was explained. Here we construct
hypergeometric functions related to the projective Schur functions
as certain tau functions of the one component BKP hierarchy. The
consideration is going along the way of
\cite{pd22},\cite{OS2},\cite{tmf}. Let us notice that a different
KP hierarchy of the root type B was constructed in the paper
\cite{KLCRM}, we do not study this case in the present paper.
\subsection{Neutral free fermions and BKP hierarchy
\cite{DJKM},\cite{DJKM'}}

Let us consider  neutral fermions $\{\phi_n,n \in Z\}$,  obeying
the following canonical anticommutation relations
\begin{equation}\label{anticomB}
[\phi_m,\phi_n]_+=(-)^m\delta_{m,-n},
\end{equation}
where $[a,b]_+=ab+ba$. In particular $\phi_0^2=1/2$.

 There are the right and the left vacuum vectors, $|0\r $ and
 $ \l 0| $ respectively, having the properties
\begin{equation}\label{vak'}
 \phi_m |0\r=0 \quad (m<0),
\qquad \l 0|\phi_m=0 \quad (m> 0)
\end{equation}
We have a right and a left Fock spaces spanned, respectively, by
the right and the left vacuum vectors and by right and left
vectors
\begin{equation}\label{Fock}
\phi_{n_1}\cdots \phi_{n_k} |0\r ,\quad \l 0|\phi_{-n_k}\cdots
\phi_{-n_1},
\end{equation}
where $k=1,2,\dots$, and where we put
\begin{equation}\label{1...k}
n_1>\cdots > n_k \ge 0
\end{equation}
 According to \cite{DJKM}, \cite{DJKM'} we introduce the
following operators (Hamiltonians) labeled by odd numbers $n \in
2Z+1$:
\begin{equation}
H_n= \frac{1}{2}\sum_{k=-\infty}^\infty (-1)^{k+1}\phi_k
\phi_{-k-n}
\end{equation}
 One can check that $H_n$ obey the following Heisenberg algebra
relations
\begin{equation}\label{HeisB}
[H_n,H_m]=\frac{n}{2}\delta_{m+n,0},
\end{equation}
where $[a,b]:=ab-ba$ is a commutator.

By (\ref{anticomB}) we obtain
\begin{equation}\label{Hphi_n}
[H_n,\phi_m]=\phi_{m-n}
\end{equation}

We also notice that
\begin{equation}\label{vacH}
H_{n}|0\r=H_{n}\phi_0 |0\r=0, \quad \l 0|H_{-n}=\l 0|\phi_0
H_{-n}=0,\quad n>0
\end{equation}

For a pair of collections of independent variables (so called BKP
higher times, whose numbers we choose to be always odd numbers)
\begin{equation}\label{todd}
{\bf t}=(t_1,t_3,t_5\dots),\quad {\bf
t}^*=(t_1^*,t_3^*,t_5^*\dots)
\end{equation}
 one sets
\begin{equation}\label{hamiltonians'} H ({\bf t} )=\sum_{n=1,3,\dots
}^{+\infty} t_n H_n  ,\quad {H }^*({\bf t} ^*)=\sum_{n=1,3,\dots
}^{+\infty} t_n^* H_{-n}
\end{equation}

It is suitable to introduce the following fermionic field, which
depends on a complex parameter $z$:
\begin{equation}\label{phiz}
\phi(z)=\sum_{-\infty}^\infty z^k\phi_k
\end{equation}

 For fermionic fields
(\ref{phiz}) the anticommutation relation (\ref{anticomB}) reads
as
\begin{equation}\label{anticomzB}
 [\phi(z),\phi(z')]_+=\delta\left(-z,z'\right)
=\sum_{n\in Z}\left(\frac{-z}{z'}\right)^n ,
\end{equation}
(the symbol $\delta$ denotes a Dirac $\delta$-function with the
defying property $\frac{1}{2\pi i}\oint
f(z)\delta\left(-z/z'\right)\frac{dz}{z}=f(-z')$).

The relation (\ref{Hphi_n}) yields
\begin{equation}\label{Hphiz}
[H_n,\phi(z)]=z^n\phi(z) ,
\end{equation}
which in turn results in
\begin{equation}\label{xitB}
 e^{H ({\bf t} )}\phi(z) e^{-H ({\bf t} )}=
\phi(z) e^{\xi({\bf t} ,z)}, \quad \xi({\bf t}
,z)=\sum_{n=1,3,\dots } z^n t_n
\end{equation}

The formulae below are called bosonization formulae
\cite{DJKM'},\cite{You}:
\begin{equation}
\l 0|\phi(z)e^{H ({\bf t} )}= \l 0|\phi_0 e^{H \left({\bf t}
-\epsilon (z^{-1})\right)} ,
\end{equation}
\begin{equation}
 \sqrt{2}\l 0|\phi_0\phi(z)e^{H ({\bf t} )}=
\l 0|e^{H \left({\bf t} -\epsilon (z^{-1})\right)} ,
\end{equation}
where
\begin{equation} \epsilon (z^{-1})=
\left(\frac{2}{z},\frac{2}{3z^3}, \frac{2}{5z^5},\dots \right)
\end{equation}
Furthermore, given $N>1$, one obtains bosonization relations:
\begin{equation}\label{bosonization}
\l 0|e^{H({\bf t}({\bf x}^N))} = 2^{\frac {N}{2}}\frac{\l 0|\phi
(\frac{1}{x_N}) \cdots\phi (\frac{1}{x_1})} {{ \Delta}({\bf x}^N)}
,\quad N \quad \textrm{even} ,
\end{equation}
\begin{equation}
\label{bosonization'} \l 0|e^{H({\bf t}({\bf x}^N))} = 2^{\frac
{N+1}{2}}\frac{\l 0|\phi_0\phi (\frac{1}{x_N}) \cdots\phi
(\frac{1}{x_1})} {{\Delta}({\bf x}^N)} ,\quad N \quad \textrm{odd}
,
\end{equation}
where
\begin{equation}\label{vand'}
\Delta({\bf x}^N)=
\frac{\prod_{i<j}^N(1-x_jx_i^{-1})}{\prod_{i<j}^N(1+x_jx_i^{-1})}=
\frac{\prod_{i<j}^N(x_i-x_j)}{\prod_{i<j}^N(x_i+x_j)}
\end{equation}

The following expression is a typical example of BKP tau-function
\begin{equation}\label{tauB}
\tau ({\bf t} )=\l 0| e^{H ({\bf t} )}g| 0 \r,\quad
g=\prod_{k=1}^{2K} \Phi_k \exp
\left(\sum_{n,m}b_{n,m}:\phi_n\phi_{-m}:\right) ,
\end{equation}
where each of $\Phi_k$ is a linear combination of neutral fermions
$\phi_k$, and ${\exists N}: b_{n,m}=0,|n-m|>N$. Symbol $:~:$ means
normal ordering, i.e. $:\phi_n\phi_{-m}:$ is $\phi_n\phi_{-m}-\l
0|\phi_n\phi_{-m}|0\r$. The numbers $b_n$ we consider satisfy
 $b_{n,m}+b_{-m,-n}=0$.\\

We remind definitions \cite{Mac} will be required in the sequel. A
set of non increasing positive integers $n_1 \ge n_2 \ge\cdots\ge
n_k\ge 0$ is called the {\em partition} of number
$n=|{\lambda}|=n_1+\dots +n_k$, and is denoted by
${\lambda}=(n_1,n_2,\dots ,n_k)$;
 if $n_k>0$, then $k$ is called the {\em length} of
the partition ${\lambda}$ and denoted by $l(\lambda)$. The number
$|{\lambda}|$ is called the weight of the partition ${\lambda}$.
Numbers $n_1,n_2,\dots$ are said to be the {\em parts} of the
partition ${\lambda}$. The partition of zero (i.e. $n_1=0$) is
denoted by ${\lambda}=0$. The set of all partitions is usually
denoted by $P$.

The {\em diagram} of a partition ${\lambda}$ (or the Young diagram
${\lambda}$) may be defined as the set of points (nodes) $(i,j)
\in Z^2$ such that $1\ge j \ge n_i$.

The conjugate of a partition ${\lambda}$ is the partition
${\lambda}'$ whose diagram is the transpose of the diagram
${\lambda}$, i.e. the diagram obtained by reflection in the main
diagonal.

There is different notation for partitions (Frobenius notation).
Suppose that the main diagonal of the diagram ${\lambda}$ consists
of $r$ nodes $(i,i)~$ $(1\le i\le r)$. Let $\alpha_i=n_i-i$ be the
number of nodes in the $i$th row of ${\lambda}$ to the right of
$(i,i)$, for $1 \le i\le r$, and let $\beta_i=n_i'-i$ be the
number of nodes in the $i$th column of ${\lambda}$ below $(i,i)$,
for $1 \le i\le r$. One has $\alpha_1>\alpha_2>\cdots >\alpha_r
\ge 0$ and $\beta_1>\beta_2>\cdots >\beta_r \ge 0$. Then we denote
the partition ${\lambda}=(n_1,n_2,\dots ,n_k)$ by
\begin{equation}
{\lambda}=(\alpha_1,\dots ,\alpha_r |\beta_1,\dots ,\beta_r
)=(\alpha |\beta )
\end{equation}

Given set of variables ${\bf t}=(t_1,t_2,t_3,\dots)$ and a given
partition $\lambda=(n_1,\dots,n_k),n_k>0$, {Schur function} is
defined as
\begin{equation}\label{Schurfunction}
s_\lambda({\bf t})=\det h_{n_i-i+j}({\bf
t})|_{i,j=1,\dots,k},\quad \sum_{k=0}^\infty z^kh_k({\bf
t}):=e^{\sum_{m=1}^\infty z^mt_m}
\end{equation}
In what follows, we shall need only those sets of variables ${\bf
t}$ where all variables with even numbers vanish: ${\bf
t}=(t_1,0,t_3,0,t_5,\dots)$. Keeping this in mind we shall write
${\bf t}=(t_1,t_3,t_5,\dots)$ as in (\ref{todd}). (Below we shall
use projective Schur functions which are defined as polynomials in
variables ${\bf t}=(t_1,t_3,t_5,\dots )$ ).

A set of strictly vanishing positive integers
${\lambda}=n_1>n_2>\cdots>n_k\ge 0$ is called the partition with
distinct parts or the {\em strict partition}. The set of all
strict partitions is denoted by $DP$.

We see that vectors (\ref{Fock}) are labeled by strict partitions.
Zero partition ${\lambda}=0$ is related to vacuum vectors $\l
0|,|0 \r$ .

If ${\lambda}=(n_1,n_2,\dots ,n_k)\in DP$ and $n_k\neq 0$, one
gets the so-called {\em double} of $\lambda$, which is the
partition ${\tilde {\lambda}}=(n_1,\dots ,n_k | n_1-1,\dots
,n_k-1) \in P$ in the Frobenius notation.

{\bf Lemma 1 \cite{You}} Let $n_1>n_2>\cdots>n_k\ge 0$, $k$ is
even. If $n_k=0$ we put  $\lambda=(n_1,n_2,\dots,n_{k-1})$, and if
$n_k >0$ we put $\lambda=(n_1,n_2,\dots,n_{k})$. Then
\begin{equation} \l 0|e^{H ({\bf t} )}\phi_{n_1}
\cdots\phi_{n_k}  |0\r
 =2^{-\frac k2}Q_{{\lambda}}\left(\frac{{\bf t}}{2}\right) ,
\end{equation}
where $Q_{{\lambda}}\left(\frac{{\bf t}}{2}\right)$ is the
polynomial in variables $t_1,t_3,t_5,\dots$, which is called a
{\em projective Schur function} and may be presented via the usual
Schur function $s_\lambda$ as
\begin{equation}\label{sQ}
2^{-\frac k2}Q_{{\lambda}}\left(\frac{{\bf t} }{2}\right)= \sqrt
{s_{{\tilde \lambda}}({\bf t} )},
\end{equation}
where $s_{ {\tilde \lambda}}({\bf t} )$ is the Schur function, and
the partition ${ {\tilde \lambda}}\in P$ is the double of the
strict partition $ {\lambda }$. Notice that
$Q_{{\lambda}}\left(\frac{{\bf
t}}{2}\right)=Q_{{\lambda}}\left(\frac{t_1}{2},
\frac{t_3}{2},\frac{t_5}{2},\dots\right)$, while $s_{ {\tilde
\lambda}}({\bf t} )=s_{ {\tilde \lambda}}(t_1,0,t_3,0,t_5,0,\dots
)$. $\Box$

We shall often use a special choice of times:  ${\bf t} ={\bf
t}({\bf x}^{N_1}),{\bf t }^* = {\bf t}^*({\bf y}^{N_2})$:
\begin{equation}\label{x/-x}
mt_m=\sum_k^{N_1} \left(x_k^m-(-x_k)^m\right),\quad
mt_m^*=\sum_k^{N_2} \left(y_k^m-(-y_k)^m\right)
\end{equation}
In case when it is not confusing we shall omit the superscripts
$N,N^*$, mainly we shall consider the case $N_1=N_2=N$ with some
$N$.

Being rewritten as a function of $x$ variables, $Q_\lambda(x)$
becomes a well-known symmetric function, which was invented by
Schur in the construction of the projective representations of the
symmetric groups \cite{Schur}.

Also we shall need the so-called {\em the
product-of-the-hooks-length of the shifted diagram} \cite{Mac},
which generalizes the notion of the factorial for the strict
partition, and which we shall denote by $H^*_\lambda$:
\begin{equation}\label{Hlambda}
H^*_\lambda=\left(\prod_{i=1}^k
n_i!\right)\prod_{i<j}\frac{n_i+n_j}{n_i-n_j} = 2^{-\frac
k2}\sqrt{H_{\tilde \lambda}},\quad \frac{1}{\sqrt{H_{\tilde
\lambda}}}=\sqrt{s_{\tilde \lambda}({\bf t}_\infty)}= 2^{-\frac
k2}Q_\lambda(\frac{{\bf t}_\infty}{2}),
\end{equation}
where $\frac{{\bf t}_\infty}{2}=(\frac 12,0,0,0,\dots)$, $n_i$ are
parts of $\lambda$, and $H_{\tilde \lambda}$ is the {\em
product-of-the-hooks-length of diagram} \cite{Mac} of the double
of the partition $\lambda $. The very last equality of
(\ref{Hlambda}) follows from (\ref{sQ}), other equalities of
(\ref{Hlambda}) are extracted from \cite{Mac}. The quantity
$|\lambda|!Q_\lambda(\frac{{\bf t}_\infty}{2})$ has a
combinatorial meaning: it counts the number of shifted standard
tableaux of a given shape, see \cite{Mac}.

\subsection{BKP tau-function of hypergeometric type and additional
symmetries} Let us consider the following set of commuting
fermionic operators
\begin{equation}\label{dopsim''}
B_k=\frac{1}{2} \sum_{n=-\infty}^\infty (-1)^{n}\phi_n \phi_{k-n}
r(n)r(n-1)\cdots r(n-k+1), \quad k=1,3,\dots  ,
\end{equation}

These operators are defined by a function $r$ which satisfies the
relation
\begin{equation} \label{Br}
r(n)=r(1-n).
\end{equation}
Therefore it is enough to define $r$ only for positive $n$.

We check that $[B_m,B_k]=0$ for each $m,k$.

It is easy to check that
\begin{equation}\label{vacB}
\l 0| B_k=0
\end{equation}

Let us note that one can rewrite operators $B_n$ in the form
\begin{equation}
B_k=-\frac{1}{4\pi \sqrt{-1}} \oint \frac{dz}{z}\phi(-z)
\left(\frac{1}{z}r\left(D\right)\right)^k\phi(z)
 ,\quad D:=z\frac{d}{dz} ,
\end{equation}
where operator $r(D)$ acts on all functions of $z$ from the right
hand side according to the rule $r(D)\cdot z^k=r(k)z^k$.

 For the
collection of independent variables ${\bf t}^*
=(t^*_1,t^*_3,\dots)$ let
\begin{equation}\label{AbetaB}
B({\bf t}^* )=\sum_{n=1,3,\dots}  t^*_nB_n
\end{equation}
For a partition ${\lambda}=(n_1,\dots ,n_k)$ and for a function of
one variable $r$, we introduce the  notation
\begin{equation}\label{Br_n}
r_{{\lambda}}=\prod_{i=1}^{k} r(1)r(2)\cdots r(n_i).
\end{equation}
We set $r _{{\bf 0}}(M)=1$.

We have\\
{\bf Lemma 2} Under conditions of the Lemma 1 the following
formula holds
\begin{equation}
\l 0|\phi_{-n_k} \cdots\phi_{-n_1}e^{-B({\bf t}^*  )}  |0\r
 = 2^{-\frac k2}
r _{{\lambda}}Q_{{\lambda}}\left(\frac{{\bf t}^* }{2}\right)
\end{equation}

Now let us consider the following tau-function (\ref{tauB}) of the
BKP hierarchy
\begin{equation}\label{tauhyp1B}
\tau _r({\bf t} ,{\bf t}^*  ):=\langle 0|e^{H ({\bf t} )}
e^{-B({\bf t}^* )}  |0\rangle
\end{equation}
Remark. Here $B_k$ generate symmetry flows for BKP hierarchy
(similar to the {\em additional symmetries of KP \cite{O}}). The
symmetries act on vacuum solution, parameters $t_k^*$ play the
role of group times.

Using Taylor expanding $e^H=1+H+\cdots$ and {\em Lemma 2} we
easily get \bprop We have the expansion:
\begin{equation}\label{tauhypB}
\tau _r({\bf t} ,{\bf t}^*  ) = 1+\sum_{{\lambda}\in
DP}2^{-l({\lambda})} r _{ \lambda }Q_{\lambda }\left(\frac{{\bf t}
}{2}\right) Q_{\lambda }\left(\frac{{\bf t}^* }{2}\right) ,
\end{equation}
 where sum is going over all
nonzero strict partitions (partitions with distinct parts $DP$).
 $l({\lambda})$ is the length of the partition ${\lambda}$.
\eprop

Obviously we get:
\begin{equation}\label{tausimtbetaB}
\tau_r ({\bf t} ,{\bf t}^* )=\tau_r ({\bf t}^* ,{\bf t} )
\end{equation}
Also for $a \in C$ tau function $\tau_r ({\bf t} ,{\bf t}^* )$
does not change if $t_m \to a^mt_m, t^*_m \to
a^{-m}t^*_m,m=1,3,\dots $. And it does not change if $t_m \to
a^mt_m,m=1,3,\dots, r(n) \to a^{-1} r(n) $.

For ${\bf t} ={\bf t}({\bf x}^{N_1}),{\bf  t }^* = {\bf t}^*({\bf
y}^{N_2})$, see (\ref{x/-x}), the sum (\ref{tauhypB}) is
restricted to the sum over partitions of the length
$l({\lambda})\le N=min(N_1,N_2)$.

Remark. Let us cite the recent paper \cite{TW} where a certain
analog of Gessel theorem was proved, concerning actually the sum
(\ref{tauhypB}) with very specific choice of $r$: $r(n)=1,n<M$ and
$r(n)=0,n\ge M$. (This case describes certain rational solutions
of BKP).

{\bf Remark}. If there exists a function $\rho(n),n \in Z$  such
that
\begin{equation}\label{rhor} r(n)=\rho(-n)\rho(n-1),
\end{equation}
 the constraint (\ref{Br}) is satisfied. Then it is easy to check that
\begin{equation}\label{rrho}
 r_{{\lambda}}=\rho^{KP}_{\bf {\tilde \lambda}}(0),
\end{equation}
the $\rho^{KP}_{\bf {\tilde \lambda}}(k)$ was used to construct
hypergeometric functions related to Schur functions in the
framework of KP theory \cite{tmf},\cite{OS2} and was defined as
\begin{equation}\label{rho^KP}
\rho^{KP}_{\bf {\tilde \lambda}}(k):=\prod_{i,j\in \bf {\tilde
\lambda}} \rho(k+j-i)
\end{equation}
In (\ref{rrho}) the product goes over all nodes of Young diagram
${\bf {\tilde \lambda}}$ and the partition ${\bf {\tilde
\lambda}}$ is the double of the strict partition $ {\lambda }\in
PD $. In this case
\begin{equation}
\l 0|\phi_{-n_k} \cdots\phi_{-n_1}e^{-B({\bf t}^*  )}  |0\r
 = 2^{-\frac k2}
r _{{\lambda}}Q_{{\lambda}}\left(\frac{{\bf t}^* }{2}\right)=
\rho^{KP}_{\bf {\tilde \lambda}}(0)\sqrt {s_{\bf {\tilde
\lambda}}({\bf t}^* )}.
\end{equation}
\begin{equation}\label{tauhypBrho}
\tau _r({\bf t} ,{\bf t}^*  ) = 1+\sum_{{\lambda}\in
DP}2^{-l({\lambda})} r _{ \lambda }Q_{\lambda }\left(\frac{{\bf t}
}{2}\right) Q_{\lambda }\left(\frac{{\bf t}^* }{2}\right) =
1+\sum_{{\lambda}\in DP}\rho^{KP}_{\bf {\tilde \lambda}}(0) \sqrt
{s_{\bf {\tilde \lambda}}({\bf t} ) s_{\bf {\tilde \lambda}}({\bf
t}^* )},
\end{equation}
 where sum is going over all
nonzero strict partitions (partitions with distinct parts $DP$),
$r(n)=\rho(-n)\rho(n-1)$, while the partition ${\tilde \lambda}\in
P$ is the double of the strict partition $ {\lambda }\in PD$. The
notation $\rho^{KP}_{\bf {\tilde \lambda}}(0)$ is given by
(\ref{rho^KP}),
 and the notation
$r _{ \lambda }$ see in (\ref{Br_n}). $l({\lambda})$ is the length
of the partition ${\lambda}$. $\Box$

\bprop

Let us consider KP tau function of hypergeometric type
$\tau_r(n,{\bf t},{\bf t}^*)$ \cite{OS2},\cite{tmf},\cite{pd22},
which we denote $\tau_r^{KP}(n,{\bf t},{\bf t}^*)$ in the present
paper, while the notation $\tau_r({\bf t},{\bf t}^*)$ we keep for
the BKP tau function (\ref{tauhyp1B}). For $r(n)=r(1-n)$ we have
\begin{equation}\label{tauBtau}
\tau_r^{KP}(0,{\bf t},{\bf t}^*)=\tau_r^2({\bf t},{\bf t}^*) ,
\end{equation}
or
\begin{equation}\label{hypQhyps}
 \left(1+\sum_{{\lambda}\in
DP}2^{-l({\lambda})} r _{ \lambda }Q_{\lambda }\left(\frac{{\bf t}
}{2}\right) Q_{\lambda }\left(\frac{{\bf t}^* }{2}\right)
\right)^2= 1+\sum_{{\lambda } \in P}r^{KP}_{\lambda }(0)s_{\lambda
}({\bf t} ) s_{\lambda }({\bf t}^* ),
\end{equation}
 where the notation $r _{ \lambda }$ in the left
hand side is due to (\ref{Br_n}), and the notation
$r^{KP}_{\lambda}(0)$ in the right hand side is defined in
(\ref{rho^KP}). The sum in the left hand side is going over all
nonzero partitions with distinct parts, while the sum in the right
hand side is going over all nonzero partitions. \eprop

Let us consider additional set of commuting operators ${\tilde
B}_k$:
\begin{equation}\label{tildedopsim'}
{\tilde B}_k=\frac{1}{2} \sum_{n=-\infty}^\infty (-1)^{n+1}\phi_n
\phi_{-k-n} {\tilde r}(n+1){\tilde r}(n+2)\cdots {\tilde r}(n+k),
\quad k=1,3,\dots  ,
\end{equation}
\begin{equation}
{\tilde B}({\bf t}^* )=\sum_{k=1,3,\dots}{\tilde B}_kt_k^*
\end{equation}
One can also rewrite these operators in the form
\begin{equation}
{\tilde B}_n=-\frac{1}{4\pi \sqrt{-1}} \oint \frac{dz}{z}\phi(-z)
\left({\tilde r}\left(D\right)z\right)^n\phi(z)
 ,\quad D:=z\frac{d}{dz} ,
\end{equation}
where operator $r(D)$ acts on all functions of $z$ from the right hand side.\\

\bprop
\begin{equation}\label{trrB}
\langle 0|e^{\tilde{B}( {\bf t}  )} e^{-B({\bf t}^*  )} |0\rangle
= \sum_{{\lambda}\in DP} (\tilde{r}r)_{ \lambda }
2^{-l({\lambda})}Q_{\lambda }(\frac{{\bf t} }{2} ) Q_{\lambda
}(\frac{{\bf t }^* }{2})
\end{equation}
\eprop

Taking ${\tilde r}r=1$ we get the useful relation:
\begin{equation}\label{orttt*B}
\exp\left(\sum_{n=1,3,5,\dots}\frac 12
nt_nt_n^*\right)=\sum_{{\lambda}\in DP}
2^{-l({\lambda})}Q_{\lambda }(\frac{{\bf t} }{2}) Q_{\lambda
}(\frac{{\bf t}^* }{2}),
\end{equation}
where the l.h.s. one gets by evaluating the vacuum expectation
value $\langle 0|e^{H( {\bf t}  )} e^{H({\bf t}^*  )} |0\rangle$
with the help of (\ref{HeisB}) and (\ref{vacH}).

Formula (\ref{orttt*B}) yields the so-called vacuum tau-function.

The relation (\ref{orttt*B}) will be of use throughout the text,
whatever symbols $t_k,t_k^*$ mean.

Formulas (\ref{dopsim''}) (\ref{phiz}) and (\ref{anticomB})
provide
\begin{equation}\label{Bphi}
[B_n,\phi(z)]=-\left(\frac 1z r(D)\right)^n\cdot \phi(z)
\end{equation}
Considering the exponents as its Taylor series we obtain
\begin{equation}\label{psixirB}
e^{B({\bf t}^* )}\phi(z)e^{-B({\bf t}^* )}=\sum_{n\in Z}\phi_n
w(n,z)=e^{-\xi_r({\bf t}^* ,z^{-1})}\cdot \phi(z),
\end{equation}
where operators
\begin{equation}\label{xirB}
\xi_r({\bf t}^* ,z^{-1})=\sum_{m=1,3,\dots}t_m\left(\frac 1z
r(D)\right)^m, \quad D=z\frac{d}{dz}
\end{equation}
act on all functions of $z$ on the right hand side, and
\begin{equation}
w(n,z)= z^n\sum_{m=0}^\infty z^{-m}h_m(-{\bf t}^*)r(n)\cdots
r(n-m) =e^{-\xi_{r}({\bf t}^* ,z^{-1})}\cdot z^n
\end{equation}
The formula (\ref{psixirB}) generalizes (\ref{xitB}). In
(\ref{psixirB}) $\xi_r$ are operators which act on $z$ variable of
$\phi(z)$.

\subsection{$H_0({ T})$, fermions $\phi({ T},z)$
and bosonization rules}

Let us introduce a set of variables $ {T}=\{ T_n,n \in Z;
T_n=-T_{-n}\}$. Put $:\phi_m\phi_n:=\phi_m\phi_n -\l 0|
\phi_m\phi_n |0\r $ (it is said that the notation $:A:$ serves for
a normal ordering of an operator $A$).
 Then we consider the operators
\begin{equation}\label{H0+'}
H_0 ({ T})=\frac 12 \sum_{n \in Z}(-)^{n+1}T_n:\phi_n \phi_{-n}:
=\sum_{n>0}(-)^{n+1}T_n\phi_n \phi_{-n}
\end{equation}
Then we see that
$e^{(-)^{n+1}T_n\phi_n\phi_{-n}}=1+(-)^n\phi_n\phi_{-n}
\left(e^{-T_n}-1\right)$, and therefore we obtain
\begin{equation}
e^{H_0 ({ T})}\phi_ne^{-H_0 ({ T})}=e^{-T_n}\phi_n
\end{equation}
For $r\neq 0$ we put
\begin{equation}\label{rT}
r(n)=e^{T_{n-1}-T_n}
\end{equation}
We see that $r(n)=r(1-n)$. Then
\begin{equation}\label{prop43'}
 e^{-H_0
({ T})}B({\bf t}^* ) e^{H_0 ({ T})}=-H^*({\bf t}^* ) .
\end{equation}

Let $r\neq 0$. It is convenient to consider the  fermionic
operators:
\begin{eqnarray}\label{kruchferm'}
\phi({ T},z):=e^{H_0 ({ T})}\phi(z)e^{-H_0 ({ T})}
=\sum_{n=-\infty}^{n=+\infty} e^{-T_n}z^n\phi_n
\end{eqnarray}

Given $N$ we use (\ref{x/-x}) to  derive the bosonization rules
below in a way similar to \cite{DJKM}.
\begin{equation}\label{mm*'}
e^{-B({\bf t^*}({\bf y}^N))}|0\r= 2^{\frac {N}{2}}\frac{\phi({
T},-y_1)\cdots\phi({ T},-y_N)|0\r } {\Delta({\bf y}^N)} ,\quad N
\quad \textrm{even},
\end{equation}
\begin{equation}\label{mm*''}
 e^{-B({\bf t^*}({\bf y}^N))}|0\r= 2^{\frac
{N+1}{2}}\frac{\phi({ T},-y_1)\cdots\phi({ T},-y_N)\phi_0|0\r }
{\Delta({\bf y}^N)} ,\quad N \quad \textrm{odd}\\
\end{equation}
Here $\Delta({\bf x}^N)$ see in (\ref{vand'}), for $N=1$ we put
$\Delta({\bf x}^N)=1$

Therefore in variables $x,y$ (see (\ref{x/-x})) one can write
 by (\ref{bosonization}),(\ref{bosonization'}):
\begin{equation}\label{fermicor'}
\tau_r({\bf t}({\bf x}^N),{\bf t^*}({\bf y}^{N^*}))=\l 0|e^{H({\bf
t}({\bf x}^{N}))} e^{-B({\bf t^*}({\bf y}^{N^*}))}|0\r
\end{equation}
\begin{equation}\label{fermicor'e-e}
= 2^{\frac{N+N^*}{2}}\frac{ \l 0|\phi(\frac{1}{x_N})\dots
\phi(\frac{1}{x_1}) \phi({ T},-y_1)\cdots\phi({ T},-y_{N^*})|0\r}
{{\Delta}({\bf x}^N)
 \Delta({\bf y}^{N^*})},\quad N+N^* \quad \textrm{even}
\end{equation}
\begin{equation}\label{fermicor'e-o}
= 2^{\frac{N+N^*+1}{2}}\frac{ \l 0|\phi_0\phi(\frac{1}{x_N})\dots
\phi(\frac{1}{x_1}) \phi({ T},-y_1)\cdots\phi({ T},-y_{N^*})|0\r}
{{\Delta}({\bf x}^N)
 \Delta({\bf y}^{N^*})},\quad N+N^* \quad \textrm{odd}
\end{equation}

By (\ref{anticomB}), (\ref{vak'}) we have
\begin{equation}\label{vaccoreven}
2^{\frac{N}{2}}\l 0|\phi (\frac{1}{x_N}) \cdots\phi
(\frac{1}{x_1})|0\r=\Delta({\bf x}^N) ,\quad 2^{\frac{N+1}{2}}\l
0|\phi_0\phi (\frac{1}{x_N}) \cdots\phi
(\frac{1}{x_1})|0\r=\Delta({\bf x}^N) ,
\end{equation}
where the first equality is for even $N$ while the second is for
odd $N$, and where we put
\begin{equation}\label{denominator}
 \Delta({\bf x}^N)=\Delta =
\prod_{i<j}^N\frac{1-x_jx_i^{-1}}{1+x_jx_i^{-1}},\quad
\frac{1-x_jx_i^{-1}}{1+x_jx_i^{-1}}=1
+2\sum_{n=1}^\infty\left(-\frac {x_j}{x_i}\right)^n
\end{equation}

At last we consider the different expression for the tau function
of hypergeometric type, which we obtain due to (\ref{prop43'}).
For non-vanishing $r$ we have
\begin{equation}\label{tau3'}
\tau_r ({\bf t} ,{\bf t}^* )= \langle 0|e^{H ({\bf t} )}
\exp\left(\sum_{-\infty}^{\infty}(-)^{n+1}T_n:\phi_n
\phi_{-n}:\right) e^{{{H }^*({\bf t}^* )}}|0\rangle
\end{equation}
\begin{equation}\label{tauhyp''}
 =1+ \sum_{{\lambda}\in DP}2^{-l({\lambda})}
e^{-T_{n_1}-T_{n_2}-\cdots -T_{n_l} }Q_{\lambda }(\frac{\bf t}{2}
) Q_{\lambda }(\frac{{\bf  t }^*}{2} ) .
\end{equation}
The sum is going over all different strict partitions
${\lambda}=(n_1,n_2,\dots,n_l),\quad l=1,2,3,... $, excluding the
partition of zero.

Tau-function (\ref{tau3'}) is linear in each $e^{T_n}$. With
respect to the BKP dynamics the times $T_n$ have a meaning of
integrals of motion.

\subsection{Expressions and linear equations for the tau-functions
$\tau_r({\bf t}({\bf x}^N),{\bf t}^* ) $, $\tau_r\left({\bf
t}({\bf x}^N),{\bf t}^*({\bf y}^{N^*}) \right) $}

It is the well-known fact that tau functions solve Hirota bilinear
equations \cite{DJKM},\cite{DJKM'}. In this section we write down
linear equations, which follow from the
 the bosonization formulae (\ref{fermicor'}).
These equations may be viewed as constraint on the totality of BKP
tau functions, or, as someone prefer, may be viewed as a certain
form of the so-called string equations.

First let us consider tau function (\ref{tauhypB}) as a function
of variables ${\bf t}({\bf x})$ (see (\ref{x/-x})) and of ${\bf
t}^* =(t_1,t_3,\dots)$.

Let $r'(n):=r(-n)$. By (\ref{xirB}) we have
\begin{equation}\label{xirxB}
\xi_{r'}({\bf t}^* ,x_i)
=\sum_{m=1,3,\dots}t_m\left(x_ir(D_i)\right)^m, \quad
D_i=x_i\frac{\partial}{\partial x_i}
\end{equation}
Let us note that $[\xi_{r'}({\bf t}^* ,x_n),\xi_{r'}({\bf t}^*
,x_m)]=0$ for all $n,m$. Below  each $e^{\xi_{r'}({\bf t}^*
,x_k)}$ means the Taylor series $1+ \xi_{r'}({\bf t}^*
,x_k)+\cdots $.

Then in cases ${\bf t} ={\bf t}({\bf x}^N)$ (using (\ref{vacB}),
(\ref{mm*'})),(\ref{vaccoreven}) and (\ref{psixirB})) we have the
following representations
\begin{equation}\label{tau+xiB}
\tau_r ({\bf t}({\bf x}^N),{\bf t}^* )=
\Delta^{-1}e^{\xi_{r'}({\bf t}^* ,x_1)}\cdots e^{\xi_{r'}({\bf
t}^* ,x_N)}\cdot \Delta ,
\end{equation}
where  each factor in the product of $\Delta$ is the notation of
the infinite series (\ref{denominator}). Therefore the action of
the pseudo-differential operators
$r(D_k),D_k=x_k\frac{\partial}{\partial x_k}$, are well defined on
$\Delta$ via the action on each monomial $\prod_i x_i^{n_i}$, the
action which is given by
\begin{equation}\label{Dkmon}
r(D_k)\cdot \prod_i x_i^{n_i}=r(n_k)\prod_i x_i^{n_i}
\end{equation}
For $N=1$ the expression (\ref{tau+xiB}) takes a simple form:
\begin{equation}\label{taux}
\tau_r ({\bf t}({x}),{\bf t}^* )=1+r(1)xh_1({\bf
t}^*)+r(1)r(2)x^2h_2({\bf t}^*)+ \cdots
\end{equation}

From (\ref{tau+xiB}) it follows that

\begin{equation}\label{linur'}
\left(\frac{\partial}{\partial t^*_m}-\sum_{i=1}^N
(x_ir(-D_{x_i}))^m \right)\cdot\left(\Delta \tau_r ({\bf t}({\bf
x}^N),{\bf t}^* )\right)=0
\end{equation}

Now let us consider the case $ {\bf t}={\bf t}({\bf x}^N), {\bf
t}^*={\bf t}^*({\bf y}^{N^*})$, see (\ref{x/-x}). We get
\begin{eqnarray}\label{repr}
\tau_r( {\bf t}({\bf x}^N), {\bf t}^*({\bf y}^{N^*}))
=\sum_{{\lambda}\in P}r_{ \lambda }2^{-\l({\lambda})} Q_{\lambda
}({\bf x}^N)Q_{\lambda }({\bf y}^{N^*})\\ \label{reprx}
=\frac{1}{\Delta({\bf x}^N)}
\prod_{i=1}^N\prod_{j=1}^{N^*}(1+y_jx_ir(D_{x_i}))\cdot
(1-y_jx_ir(D_{x_i}))^{-1}\cdot \Delta({\bf x}^N)\\ \label{repry}
=\frac{1}{\Delta({\bf y}^{N^*})}
\prod_{i=1}^{N}\prod_{j=1}^{N^*}(1+x_iy_jr(D_{y_j}))\cdot
(1-x_iy_jr(D_{y_j}))^{-1}\cdot \Delta({\bf y}^{N^*}) ,
\end{eqnarray}
where $(1-y_jx_ir(D_{x_i}))^{-1}$ and $(1-x_iy_jr(D_{y_j}))^{-1}$
are  formal series $1+y_jx_ir(D_{x_i})+\cdots $ and
$1+x_iy_jr(D_{y_j})+\cdots $ respectively.

Looking at (\ref{reprx}) ,(\ref{repry}) one derives the following
system of linear equations
\begin{equation}\label{linuryx}
\left(D_{y_j}-2\sum_{i=1}^N\frac{y_jx_ir(D_{x_i})}
{1-\left(y_jx_ir(D_{x_i})\right)^2} \right)\cdot\left(
\Delta\left({\bf x}^N\right) \tau_r\left({\bf t}\left({\bf
x}^N\right), {\bf t}^*\left({\bf
y}^{N^*}\right)\right)\right)=0,\quad j=1,\dots,N^*,
\end{equation}
\begin{equation}\label{linurxy}
\left(D_{x_i}-2\sum_{j=1}^{N^*}\frac{x_iy_jr(D_{y_j})}
{1-\left(x_iy_jr(D_{y_j})\right)^2} \right)\cdot\left(
\Delta\left({\bf y}^{N^*}\right) \tau_r\left({\bf t}\left({\bf
x}^N\right), {\bf t}^*\left({\bf
y}^{N^*}\right)\right)\right)=0,\quad i=1,\dots,N ,
\end{equation}
where $\left(1-\left(y_jx_ir(D_{x_i})\right)^2\right)^{-1}$ and
$\left(1-\left(x_iy_jr(D_{y_j})\right)^2\right)^{-1}$ are the
formal series $1+\left(y_jx_ir(D_{x_i})\right)^2+\cdots $ and
$1+\left(x_iy_jr(D_{y_j})\right)^2+\cdots $ respectively.

 Also for the tau-function written in variables
${\bf x}^N,{\bf y}^{N^*}$ (see (\ref{x/-x})) we have
\begin{equation}\label{linur2'}
\frac{1}{\Delta} \sum_{i=1}^N D_{x_i} {\Delta}\tau ({\bf t}({\bf
x}^N),{ T}, {\bf t}^{*}({\bf y}^{(N^*)})) =\frac{1}{\Delta^*}
 \sum_{i=1}^{N^*}
\left(\frac{1}{y_i}D_{y_i}y_i\right) {\Delta^*} \tau ({\bf t}({\bf
x}^N),{\bf T},{\bf t}^{*}({\bf y}^{N^*})),
\end{equation}
where $\Delta =\Delta ({\bf x}^N)$ and $\Delta^*=\Delta({\bf
y}^{N^*})$. This formula may be obtained by the insertion of the
fermionic operator $ res_z:\phi (-z)z\frac{d}{dz}\phi(z):$ inside
the vacuum expectation value (\ref{tau3'}) (like it was done in
\cite{nl64}). These formulae can be also written in terms of
higher BKP times, with the help of vertex operator action, see the
subsection ``Vertex operator action''. Then the relation
(\ref{linur'}) is the infinitesimal version of
(\ref{stringvert1'}), while the relation (\ref{linur2'}) is the
infinitesimal version of (\ref{stringvert2'}).

\subsection{Vacuum expectation as a scalar product. Symmetric
function theory consideration}

It is well-known fact in the theory of symmetric functions, that
there exists the scalar product, where the projective Schur
functions are orthogonal
\begin{equation}\label{ortSchurB}
< Q_\mu,Q_\lambda>  =2^{l(\lambda)}\delta_{\mu,\lambda}\quad ,
\end{equation}
see \cite{Mac}, Chapter III. Let us note that projective Schur
functions form a basis, see \cite{Mac} for details.

One may obtain the following realization of this scalar product if
functions $Q_\lambda$ are written as functions of variables ${\bf
t}= (t_1,t_3,\dots)$ (sometimes below it is convenient to use the
letter $\gamma$ instead of the $t$). For functions
$Q_\lambda(\frac{\gamma}{2})$ (moreover for any functions
$f=f(\frac{\gamma}{2}),g=g(\frac{\gamma}{2})$) we have the
following realization of the scalar product:
\begin{equation}\label{scalarproductB}
<Q_\mu,Q_\lambda>=\left(Q_\mu(\frac{\tilde \partial}{2} )\cdot
Q_\lambda(\frac{\gamma}{2}) \right)|_{{\gamma} =0},\quad
<f,g>=\left(f(\frac{\tilde \partial}{2} )\cdot g(\frac{\gamma}{2})
\right)|_{{\gamma} =0} ,
\end{equation}
where
\begin{equation}\label{derivativesB}
\frac{\gamma}{2}=(\frac{\gamma_1}{2},\frac{\gamma_3}{2},
\frac{\gamma_5}{2},\dots),\quad
 \frac{\tilde \partial}{2} =(\partial_{\gamma_1},\frac 13
\partial_{\gamma_3},\dots,\frac {1}{2n-1}
\partial_{\gamma_{2n-1}},\dots)
\end{equation}
In particular we have $<\gamma_n,\gamma_m>
=<2\frac{\gamma_n}{2},2\frac{\gamma_m}{2}>=\frac 2n \delta_{n,m}$.

Let us deform the relation (\ref{ortSchurB}) as follows
\begin{equation}\label{rortSchurB}
< Q_\mu,Q_\lambda >  _{r}=2^{l(\lambda)}r_\lambda
\delta_{\mu,\lambda}
\end{equation}
By this formula and by (\ref{orttt*B}) we obtain that $\tau_r$ of
(\ref{tauhypB}) is the following scalar product of functions of
variables ${\gamma}$
\begin{equation}\label{tauscalB}
\tau_r({\bf t} ,{\bf t^*} )=< e^{\sum_{m=1,3,\dots} \frac 12 mt_m
\gamma_m},e^{\sum_{m=1,3,\dots}\frac 12 mt_m^*\gamma_m}> _{r}\quad
,
\end{equation}
notice that in (\ref{tauscalB}) the variables ${\bf t} ,{\bf t^*}
$ play the role of parameters

Now let us show that scalar product (\ref{scalarproductB}) is
equal to the following vacuum expectation value: \bprop For
functions $f=f(\frac{\gamma}{2}),g=g(\frac{\gamma}{2})$
\begin{equation}\label{vacscalB}
< f,g >  =\l 0|f(\frac{\bf {H}}{2})g(\frac{\bf H^*}{2})|0\r ,
\end{equation}
where
\begin{equation}\label{tildeHH*B}
\frac{\bf
H}{2}=\left(\frac{H_1}{1},\frac{H_3}{3},\dots,\frac{H_{2n-1}}{2n-1}
,\dots \right),\quad \frac{\bf
H^*}{2}=\left(\frac{H_{-1}}{1},\frac{H_{-3}}{3},\dots,
\frac{H_{-2n+1}}{2n-1},\dots\right)
\end{equation}
\eprop

The proof follows from the comparing of the formulae
(\ref{HeisB}),(\ref{vacH}) with (\ref{scalarproductB}).

Also we have \bprop For functions
$f=f(\frac{\gamma}{2}),g=g(\frac{\gamma}{2})$
\begin{equation}\label{vacscalB}
< f,g>  _{r}=\l 0|f(\frac{\bf { H}}{2})g(\frac{\bf B}{2})|0\r ,
\end{equation}
where
\begin{equation}\label{tildeHH*B}
\frac{{\bf
H}}{2}=\left(\frac{H_1}{1},\frac{H_3}{3},\dots,\frac{H_{2n-1}}{2n-1},
\dots\right),\quad \frac{\bf
B}{2}=\left(B_1,\frac{B_3}{3},\dots,\frac{B_{2m-1}}{2m-1},\dots\right)
\end{equation}
\eprop

\subsection{Hypergeometric functions}
 Let all parameters $b_k$ be not equal to negative integers.
 Denote $(1,0,0,0,\dots)$ by ${\bf t}_\infty$. Let
\begin{equation}\label{E3'}
r(n)= \frac{\prod_{i=1}^p
(a_i+n-1)}{\prod_{i=1}^s(b_i+n-1)},n>0,\qquad {\bf t}^*={\bf
t}_\infty
\end{equation}
We have (see (\ref{tauhypB}),(\ref{Hlambda})) :
\begin{equation}
 \tau_r ({\bf t} ,{\bf t}^* )=
 1+\sum_{{\lambda}\in DP}2^{-l({\lambda})}
 Q_{{\lambda}}\left(\frac{{\bf t} }{2}\right)\frac{1}{H^*_\lambda}
\frac{\prod_{k=1}^{p}(a_k)_{\lambda}  }
{\prod_{k=1}^{s}(b_k)_{\lambda} } ,  \label{tauBttE3'}
\end{equation}
where the notation $(a)_{{\lambda}} $ is a version of Pochhammer
Symbol attached to a partition:
\begin{equation}\label{(a)_n }
(a)_{{\lambda}}:=\prod_{i=1}^{l(\lambda)} (a)_{n_i},\quad
(a)_{n_i}:=a(a+1)\cdots (a+n_i-1)
\end{equation}

If we take ${\bf t}=(2x,\frac{2x^3}{3},\frac{2x^5}{5},\dots)$, the
sum is going over partitions of the length one, in this case
$Q_{(n)}=2x^n,n>0,Q_0=1$, $H^*_{(n)}=n!$, and we
 obtain the ordinary generalized hypergeometric function  of one variable
\begin{equation}\label{onevarE3'}
\tau_r ({\bf t} ,{\bf t}^* )=\sum_{n=0}^\infty
\frac{\prod_{k=1}^{p}(a_k)_{n} } {\prod_{k=1}^{s}(b_k)_{n} }\frac
{x^n}{n!} ={}_{p}F_s(a_1,\dots,a_p;b_1,\dots,b_s; x)
\end{equation}

Below we consider hypergeometric series related to Schur
Q-functions with the following property: its square is a certain
hypergeometric function related to the Schur functions $s_\lambda$
\cite{pd22} via the Proposition 2, see (\ref{hypQhyps}). To
achieve this property one should take a rational function $r$
solving $r(n)=r(1-n)$.
 Let all parameters $\beta_k$ be non
half integers. We take
\begin{equation}
r(n)=\frac{\prod_{k=1}^{p}\left((n-\frac 12)^2-\alpha^2_k \right)}
{\prod_{k=1}^{s}\left((n-\frac 12)^2-\beta^2_k \right)},\quad
\rho(n)=\sqrt{-1}(\alpha -\frac 12 -n)  \label{op1''}
\end{equation}
We obtain the following hypergeometric function
\begin{equation}\label{beskshurB}
 \tau_r ({\bf t} ,{\bf t}^* ) = 1+\sum_{{\lambda}\in DP}\frac
{2^{-l({\lambda})}} {H^*_\lambda }
\frac{\prod_{k=1}^{p}(\alpha_k+\frac12)_{\lambda}
(-\alpha_k+\frac12 )_{\lambda}  } {\prod_{k=1}^{s}(\beta_k+\frac12
)_{\lambda} (-\beta_k+ \frac12 )_{\lambda} }
Q_{\lambda}\left(\frac{{\bf t} }{2}\right).
\end{equation}

\subsection{Integral representations of the scalar product $<,>_r$}

We need a function $\mu_r$ of one variable with the following
properties:
\begin{equation}\label{domainGammaB}
\int \int_\Gamma \mu_r (zz^*)z^n dzdz^*=\int \int_\Gamma
\mu_r(zz^*) {z^*}^n dzdz^*=\delta_{n,0}
\end{equation}
together with the relation
\begin{equation}\label{intzz^*B}
\int \int_\Gamma \mu_r (zz^*)z^n{z^*}^m
dzdz^*=2\delta_{n,m}r(1)r(2)\cdots r(n)
\end{equation}

Remark. The way to find an appropriate $\mu_r$ is to solve the
equation
\begin{equation}\label{stringfrB}
\left(z^*-\frac 1z r(-D)\right)\cdot \mu_r(zz^*)=0,\quad
D=z\frac{d}{dz},\quad r(k)=r(1-k)
\end{equation}
and to choose the integration domain $\Gamma$ should be chosen in
such a way that the operator $\frac 1z r(D),D=z\frac{d}{dz}$ is
conjugated to the operator $\frac 1z r(-D)$. $\Box $

Using
\begin{equation}\label{><}
\frac {1}{M!}\int \cdots \int \phi(z_M)\cdots \phi(z_1)|0\r \l 0|
\phi(z_1^*)\cdots \phi(z_M^*)\prod_{i=1}^M\mu_r
(z_iz_i^*)dz_idz_i^*=\sum_{\lambda \in DP ,l(\lambda)\le M}
2^{l(\lambda)} |\lambda \r r_\lambda\l \lambda |
\end{equation}
we obtain for partitions $\lambda,\mu$ (both partitions have
length $l(\lambda),l(\mu)\le M$)
\begin{equation}\label{QQ}
\frac {1}{M!}\int \cdots \int \Delta (z)\Delta
(z^*)Q_\lambda(z)Q_\mu(z^*) \prod_{k=1}^M
\mu_r(z_k{z^*}_k)dz_kdz^*_k=2^{l(\lambda)}r_\lambda
\delta_{\lambda\mu}=<Q_\lambda,Q_\mu>_r \quad ,
\end{equation}
where
\begin{equation}\label{vandzz^*B}
\Delta (z)=\prod_{i<j}^M\frac{(z_i-z_j)}{(z_i+z_j)},\quad \Delta
(z^*)=\prod_{i<j}^M\frac{(z_i^*-z_j^*)}{(z_i^*+z_j^*)}
\end{equation}
With the help of equalities
\begin{equation}\label{Schurdeczz^*B}
e^{\sum_{n=1,3,\dots}^\infty \sum_{k=1}^M z^n_kt_n}=\sum_{\lambda
\in DP ,l(\lambda)\le M} 2^{-l(\lambda)}Q_\lambda({\bf z}^M)
Q_\lambda({\bf t} ),
\end{equation}
\begin{equation}\label{Schurdeczz^*B'}
 e^{\sum_{n=1,3,\dots}^\infty \sum_{k=1}^M
{z^*}^n_kt_n^*}=\sum_{\lambda \in DP ,l(\lambda)\le M}
2^{-l(\lambda)}Q_\lambda({\bf z^*}^M) Q_\lambda({\bf t^*} )
\end{equation}
we evaluate the integral
\begin{equation}\label{intSchurB}
I_r (M,{\bf t} ,{\bf t^*} )=\frac {1}{M!}\int \cdots \int \Delta
(z)\Delta (z^*)\prod_{k=1}^M e^{\sum_{n=1,3,\dots}^\infty
\left(z^n_kt_n+{z^*_k}^nt^*_n\right)}\mu_r(z_k{z^*}_k)dz_kdz^*_k
\end{equation}
 We finely obtain
\begin{equation}\label{inttauB}
I_r (M,{\bf t} ,{\bf t^*} )=1+\sum_{\lambda \in DP ,l(\lambda)\le
M}2^{-l({\lambda})} r _{ \lambda }Q_{\lambda }\left(\frac{{\bf t}
}{2}\right) Q_{\lambda }\left(\frac{{\bf t}^* }{2}\right)
\end{equation}
The restriction $l(\lambda)\le M$ makes the difference between the
r.h.s. of (\ref{inttauB}) and $\tau_r({\bf t},{\bf t}^*)$. However
in case at least one of the sets ${\bf t},{\bf t}^*$ has the form
of (\ref{x/-x}) with $N$ or $N'$ no more then $M$, the integral
$I_r (M,{\bf t} ,{\bf t^*} )$ is the BKP tau function
$\tau_r({\bf t},{\bf t}^*)$.
 In this case I would like to interpret the integral $I_r (M,{\bf t} ,{\bf
t^*} )$ as an analog of Borel sum for the divergent series $\tau_r
$, similar we do in \cite{OH}.

It may be interesting to apply this treatment to matrix models
\cite{Moe} and to statistical models where partition functions
reduce to the similar type of integrals, see \cite{Kostov},
\cite{LS} for examples of similar integrals.

\subsection{Pfaffian formulae for the tau-functions
$\tau_r({\bf t}({\bf x}),{\bf t}^* ) $, $\tau_r\left({\bf t}({\bf
x}),{\bf t}^*({\bf y}) \right) $} First let us remind that a
Pfaffian of a $2N$ by $2N$ skew symmetric matrix $S$ is defined as
the square root of its determinant. We shall denote the Pfaffian
as $\textrm{Pfaff}(S)$,

Now let us enumerate all neutral fermions in (\ref{fermicor'})
from the left to the right as $\Phi_k,k=1,2,\dots ,2N$ as it is
written here
\begin{equation}
 \l 0|\phi(\frac{1}{x_N})\dots
\phi(\frac{1}{x_1}) \phi({\bf T},-y_1)\cdots \phi({\bf
T},-y_N)|0\r= \l 0|\Phi_1\Phi_2 \cdots \Phi_{2N-1}\Phi_{2N}|0\r
\end{equation}
$\phi(\frac{1}{x_N})=\Phi_1,\dots ,\phi({\bf T},-y_N)=\Phi_{2N}$.
With the help of Wick theorem (see \cite{DJKM'} for instance) we
obtain
\begin{equation}
 \l 0|\Phi_1\Phi_2 \cdots \Phi_{2N-1}\Phi_{2N}|0\r
=\textrm{Pfaff}(S),  \quad S_{km}= \l 0|\Phi_k \Phi_m |0\r
\end{equation}
The matrix $S_{nm}$ consists of the following four blocks, $N$ by
$N$ each:
\begin{equation}\label{Qkm}
S_{km}=\frac 12 \frac{x_m-x_k}{x_m+x_k},\quad 1 \le k,m\le
N,\qquad S_{km}=\frac 12 \frac{y_m-y_k}{y_m+y_k}, \quad N+1 \le
k,m\le 2N,
\end{equation}
and
\begin{equation}\label{Qkmxy}
S_{km}=-S_{mk}=\sum_{n=0}^\infty e^{-T_n}x_k^ny_m^n ,\quad 1\le
k\le N,\quad N+1 \le m\le 2N
\end{equation}
where $e^{T_{n-1}-T_n}=r(n)$. Thus in the case ${\bf t}={\bf
t}(x),{\bf t}^*={\bf t}^*(y)$, we obtain the Pfaffian formula for
(\ref{tauhypB})

\bprop For $r\neq 0$ and for ${\bf t}({\bf x}^N), {\bf t^*}({\bf
y}^N)$ defined by (\ref{x/-x}) we have
\begin{equation}\label{det2B}
\tau_r ({\bf t}({\bf x}^N),{\bf t}^*({\bf y}^N))=
\frac{\textrm{Pfaff}(S) } {{\Delta}({\bf x}^N)\Delta({\bf y}^N)},
\end{equation}
 where for the notation
 ${\Delta}({\bf x}^N)$  see (\ref{denominator}). The matrix
 $S_{km}$ is defined by
 (\ref{Qkm})-(\ref{Qkmxy}).
\eprop

\bprop For $N$ even
\begin{equation}\label{det1Beven}
\tau_r({\bf t}({\bf x}^N),{\bf t}^*)=\frac{ \textrm{Pfaff}\left(R
\right)_{i,k=1}^{N}}{ \Delta({\bf x}^N)},\quad R_{ik}=\frac 12
\frac{x_i-x_k}{x_i+x_k}\tau_r({\bf t}(x_i,x_k),{\bf t}^*),\quad
t_m(x_i,x_k)=\frac{2x^m_i}{m}+\frac{2x^m_k}{m}
\end{equation}
where
\begin{equation}\label{tauxy}
\frac 12 \frac{x_i-x_k}{x_i+x_k}\tau_r({\bf t}(x_i,x_k),{\bf
t}^*)=\l 0| \phi(\frac{1}{x_i})\phi(\frac{1}{x_k}) e^{-B({\bf
t}^*)}|0\r
\end{equation}
\begin{equation}\label{tauxy'}
= \frac 12 \sum_{n_k>n_i\ge
0}(x_k^{n_k}x_i^{n_i}-x_k^{n_i}x_i^{n_k})r_{(n_k,n_i)}
Q_{(n_k,n_i)}(\frac{{\bf t}^*}{2})
\end{equation}

For $N$ odd the formula is almost the same, provided the
additional $N+1$th row and line are added to matrix $R$:
\begin{equation}\label{det1Beven}
\tau_r({\bf t}({\bf x}^N),{\bf t}^*)=\frac{ \textrm{Pfaff}\left(R
\right)_{i,k=1}^{N+1}}{ \Delta({\bf x}^N)},\quad
R_{i,N+1}=-R_{N+1,i}= \tau_r({\bf t}(x_i),{\bf t}^*),\quad
t_m(x_i)=\frac{2x^m_i}{m}
\end{equation}
(where for explicit expression of $\tau_r({\bf t}(x_i),{\bf t}^*)$
see (\ref{taux})). \eprop

\subsection{The vertex operator action. Linear equations II}
Vertex operators $Z(z)$  act on the space $C[t_1,t_3,t_5 \dots ]$
of polynomials in infinitely many variables, and are defined by
the formulae:
\begin{equation}\label{vertex'}
Z(z)=e^{\xi({\bf t} ,z)}e^{-2\xi({\tilde
\partial} ,z^{-1})}, \quad \xi({\bf
t} ,z)=zt_1+z^3t_3+z^5t_5+\cdots,\quad {\tilde \partial}:
=(\frac{\partial}{\partial t_1}, \frac{1}{3}\frac{\partial}
{\partial t_3},\frac{1}{5}\frac{\partial}{\partial t_5},\dots)
\end{equation}

 Let us introduce the operators (the generators of additional symmetries)
 which act on functions of ${\bf t}$ variables:
\begin{equation}
\Omega_r({\bf t}^*):=- \frac{1}{2\pi \sqrt{-1}} \lim _{\epsilon
\to 0} \oint Z(-z+\epsilon) \xi_r({\bf t}^* ,z,D)Z (z)
\frac{dz}{z}
\end{equation}
(here parameters $t^*_k$ play the role of group times).
 For instance
\begin{equation}
\Omega_{r=1}({\bf t}^*) =2\sum_{n>0} (2n+1)t_{2n+1}t^*_{2n+1}
\end{equation}
We also consider
\begin{equation}
Z_{nn}=- \frac{1}{4\pi^2} \oint \frac {z^n}{{(-z^*)}^n}Z (-z^*)Z
(z) \frac{dzdz^*}{zz^*}
\end{equation}

Having in mind (\ref{rT}) it is convenient to have the second
notation for our tau function: $ \tau({\bf t} ,T,{\bf
t}^*):=\tau_r ({\bf t} ,{\bf t}^* )$. \bprop We have shift
argument formulae
\begin{equation}\label{stringvert1'}
e^{\Omega_r(\gamma)}\cdot \tau_r ({\bf t} ,{\bf t}^* )= \tau_r
({\bf t} ,{\bf t}^*  +\g)
\end{equation}
We also have
\begin{equation}\label{stringvert2'}
e^{\sum_{-\infty}^{\infty}\g_n Z_{nn}}\tau  ({\bf t},{ T},{\bf
t}^*)= \tau ({\bf t},{ T}+\g,{\bf t}^*),
\end{equation}
In particular
\begin{equation}
e^{\Omega_r({\bf t}^* )}\cdot 1= \tau_r ({\bf t} ,{\bf t}^* ) ,
\end{equation}
\begin{equation}
 e^{\sum_{-\infty}^{\infty}T_n Z_{nn}}
\exp \left(2\sum_{n=1,3,\dots} nt_nt^*_n\right)= \tau  ({\bf
t},T,{\bf t}^*).
\end{equation}
\eprop

\section*{Acknowledgements}
The paper was mainly done when the author visited Kyoto university
in 1999-2001. I am grateful to A.Kirillov, J.J.C.Nimmo,
I.Loutsenko and most of all to T. Shiota for discussions. I thank
grant RFBR 02-02-17382a. I also thank T. Shiota who recently
pointed me the paper \cite{TW}.

\end{document}